\documentclass{aa}
\usepackage{natbib}
\usepackage[utf8]{inputenc}
\usepackage{txfonts}
\usepackage{graphicx}
\usepackage[normalem]{ulem}

\usepackage{rotating}

\usepackage{amssymb}
\usepackage{amsmath}
\usepackage{color, colortbl}
\definecolor{pad}{rgb}{0.77,0.07,0.77}

\begin{document}

\title{Galactic and stellar perturbations of long-period comet motion -- Practical considerations}

\author{Piotr A. Dybczyński\thanks{\email{dybol@amu.edu.pl}} and {Sławomir Breiter\thanks{\email{breiter@amu.edu.pl} }}}

\institute{Astronomical Observatory Institute, Faculty of Physics, Adam Mickiewicz University, Słoneczna 36, Poznań, Poland
}

\date{Received \today; accepted xxxx}

  \abstract
   {Thanks to our expanding knowledge of the Galactic and stellar neighborhood of the Solar System, modern long-period comet motion studies must take into account both stellar perturbations and the overall Galactic potential.}
   {Our aim is to propose algorithms and methods that  aid in performing numerical integrations of   equations of motion for a small body of the Solar System that are much faster and with greater precision.}
   {We propose a new formulation of the equations of motion formulated in the Solar System barycentric frame, but one that accurately accounts for the differential perturbations caused by the Galactic potential. To make certain these equations are applied effectively, we provide numerical ephemerides of the Galactic positions of the Sun and a set of potential stellar perturbers. }
   {The proposed methods raise the precision by several orders of magnitude and, simultaneously, greatly reduce the necessary CPU time. The application of this approach is presented with the example of a detailed dynamical study of the past motion of comet C/2015~XY1. }
   {}
\keywords{comets: general -- comets individual:C/2015~XY1 -- methods: analytical -- methods: numerical -- stars: general -- Galaxy: general}

\titlerunning{Galactic and stellar perturbations in long period comet motion.}

\authorrunning{P.A.Dybczyński and S.Breiter}

\maketitle

\section{Introduction}

Long-period comets (LPCs) travel on near-parabolic orbits with  huge heliocentric distances ahead and behind them. When they are far from the Sun, they are gravitationally influenced by passing stars and by the tidal action of the overall Galactic potential. While the importance of stellar perturbations in their motion were recognized long ago \citep{opik:1932,oort:1950}, the differential influence of the Galaxy as a whole started to be investigated at the end of the 20th century. Among first papers presenting the practical way to include Galactic perturbations in LPCs dynamics we mention that by Julia Heisler and Scott Tremaine (\citeyear{heis-trem:1986}). They derived rather simple formulae accounting for the Galactic disk tidal action, which they recognized to be dominating. The problem of LPC motion under the Galactic disk influence can be approximated as an integrable one if the mean anomaly is removed by a perturbation method. Such an averaged solution was studied by a number of authors seeking its qualitative properties and approximate long-term solution. For more details, see, for example, \cite{matese-w:1992}, \cite{breiter-dyb:1996}, or a review paper by \cite{fouchard-f-m-v:2005} and references therein.

Adding the Galactic central bulge influence, which was first proposed by \cite{levison-d-d:2001}, leads to a qualitatively different problem that remains non-integrable even after the removal of the short-period terms. Thus, even the secular dynamics of LPCs requires numerical integration and only basic facts about the equilibria of this model have been reported \citep{BreFouRat:2008}.

In a series of papers, Królikowska and Dybczyński (e.g., Królikowska \& Dybczyński \citeyear{kroli-dyb:2020} and references therein), the authors investigate past and future motion of a large sample of the observed LPCs, taking into account both the Galactic and stellar perturbations based upon the available stellar data and the Galaxy potential models. In the Hipparcos era, there was no star identified that would significantly perturbing LPC motion and, additionally, a simplified Galactic potential that only accounted for the tidal effect of the Galactic disk was widely used.

Our knowledge of the Galactic and stellar neighborhood of our planetary system has expanded significantly in recent years.  Several models of the Galactic gravitational potential have been proposed; for instance, \cite{bajkova-bobylev:2017} present a review. The number of recognized stars potentially perturbing LPC motion has also significantly increased, particularly thanks to the {\it Gaia} mission results \citep{Gaia_mission:2016}. Based on the {\it Gaia} DR2 catalog \citep{Gaia-DR2:2018}, the StePPeD database of potential perturbers was presented by \cite{rita-pad-magda:2020}. It consists of about 650 stars or stellar systems. Following the publication of the {\it Gaia} EDR3 data, StePPeD database  is undergoing a substantial revision and the results  are available in part.

Using the modern Galactic potential model proposed by \cite{irrgang_et_al:2013}, it was shown by \cite{dyb-berski:2015} how a star trajectory could be calculated by means of the direct numerical integration in the galactocentric rectangular frame. This method was recently used to find the first stars that significantly perturb the LPC motion \citep{First-stars:2020}. Over the course of that work, it appeared that calculating LPC motion together with stars and under the overall Galactic potential (as the $N$-body problem with $N$ of the order of 400) encounters two serious problems, making the task very difficult.

The first problem comes from the fact that in the galactocentric frame, the positions of a comet and the Sun can be almost identical up to the first ten significant digits. This loss in precision leads to erroneous results in many cases, with regard to the long-term numerical integrations of the LPC motion. The only remedy is to switch to a quadruple precision, but this drastically worsens the second problem, namely, unacceptably time-consuming calculations. The $N$-body code with hundreds of interacting bodies requires hours to complete calculations for a single LPC, whereas we want to estimate the uncertainty of our results by means of the Monte Carlo sampling, that is, replacing a comet or a star with thousands of its clones drawn from the respective covariance matrices. We found this task nearly impossible, and this was the main impulse to develop alternative methods and algorithms described in this manuscript.

In Sect. \ref{differential}, we describe a new way of formulating the equations of motion  of a small body in the Solar System in a heliocentric frame, while maintaining the Galactic tides force intact. Section \ref{sect:galstel}  shows the method for including stellar perturbations in this dynamical model. The next two sections describe the overall picture of the Galactic and stellar perturbations on LPC motion obtained with the proposed methods and present a detailed example of a comet dynamics investigation available with our approach. Section \ref{sec:cloning} shows how effectively we can estimate the uncertainty of the above-mentioned results  and describes the difficulties arising from continuously  imprecise stellar data in some cases. Our summary and conclusions are given in Sect. \ref{sec:conclusions}.

\section{Exact differential formulae for  Galactic tides in the heliocentric frame}\label{differential}

In this paper, we use the term "galactocentric frame" for the reference frame whose origin is placed in the center of the Galaxy, whereas the axes are parallel to those of the standard Galactic heliocentric frame (with the same directions). This means that in the such defined galactocentric frame, the $OX$ axis is directed opposite to the Sun, the $OY$ is directed in accordance to the Sun's rotation around the Galactic center, and the $OZ$ axis is directed to the north Galactic pole, which makes this system a right-handed one. For the details and numerical parameters of the frame orientation and translation, we refer to \cite{dyb-berski:2015}.

We describe the galactocentric coordinates of the Sun as:\ \begin{equation}\label{f1}
\vec{R}_\odot  = \left(X_\odot, Y_\odot, Z_\odot \right)^\mathrm{T},
\end{equation}
and the galactocentric coordinates of a comet (or another point mass, e.g., a star) as
\begin{equation}\label{f2}
\vec{R}  = \left(X , Y , Z \right)^\mathrm{T}.
\end{equation}
If we denote the heliocentric coordinates of a comet as:
\begin{equation}\label{f3}
\vec{r}  = \left(x,y,z \right)^\mathrm{T},
\end{equation}
we have:
\begin{equation}\label{f4}
\vec{R} = \vec{R}_\odot + \vec{r}.
\end{equation}

The gravitational potential of the Galaxy $\Phi(\rho,Z)$ considered here is based on Model I from \citet{irrgang_et_al:2013}, with parameters listed in Table \ref{tab:Model-I-parameters}, see also \cite{dyb-berski:2015}. It is a revised version of the \cite{allen-santillan:1991} potential and is expressed as the sum of three components: a central bulge component $\Phi_{b}(R)$ (spherically symmetric),
an axisymmetric disk $\Phi_{d}(\rho,Z),$ and a massive spherical Galactic
halo $\Phi_{h}(R)$ (dark matter included). Thus:
\begin{equation}
\Phi(\rho,Z)=\Phi_{b}(R)+\Phi_{d}(\rho,Z)+\Phi_{h}(R)\label{eq:global_pot},
\end{equation}
where $(\rho,\varphi,Z)$ are the galactocentric cylindrical coordinates, $\rho =\sqrt{X^{2}+Y^{2}}$,
and $R=\sqrt{\rho^{2}+Z^{2}}$ is a galactocentric spherical radius introduced by Eq. (\ref{f2}). In a very recent paper by \cite{Bovy:2021} this model of a Galactic potential is the only one which is shown to be in agreement with the Solar System's galactocentric acceleration deduced from the {\it Gaia} EDR3 data \citep{GaiaEDR3-acceleration:2021}.  It should be noted that the formulae derived below can be applied to other particular Galactic potential models, with different numerical constants replacing those from Table \ref{tab:Model-I-parameters}.

\begin{table}
        \caption{\label{tab:Model-I-parameters}Galactic potential numerical parameters for model I from \citet{irrgang_et_al:2013}}

        \centering{} 
        \begin{tabular}{ll}
                \hline
                Parameter  & Value\tabularnewline
                \hline
                Galactic bulge mass $M_{b}$  & 9.51$\times10^{9}\textrm{M}_{\sun}$\tabularnewline
                Galactic disk mass $M_{d}$  & 66.4$\times10^{9}\textrm{M}_{\sun}$\tabularnewline
                Galactic halo mass $M_{h}$  & 23.7$\times10^{9}\textrm{M}_{\sun}$\tabularnewline
                bulge characteristic distance $b_{b}$  & 230 pc\tabularnewline
                disk characteristic distance $a_{d}$  & 4220 pc\tabularnewline
                disk characteristic distance $b_{d}$  & 292 pc\tabularnewline
                halo characteristic distance $a_{h}$  & 2562 pc\tabularnewline
                \hline
        \end{tabular}
\end{table}

For the Galactic bulge and disk components, the formulae are given in the form proposed by \cite{miyamoto-nagai:1975}:
\begin{equation}
\Phi_{b}(R)=\frac{-GM_{b}}{\sqrt{R^{2}+b_{b}^{2}}}=\Phi_{b}(X,Y,Z)=\frac{-GM_{b}}{\sqrt{X^{2}+Y^{2}+Z^{2}+b_{b}^{2}}},
\end{equation}
\begin{align}
\Phi_{d}(\rho,Z) &=\frac{-GM_{d}}{\sqrt{\rho^{2}+\left(a_{d}+\sqrt{Z^{2}+b_{d}^{2}}\right)^{2}}}\nonumber\\
&=\frac{-GM_{d}}{\sqrt{X^{2}+Y^{2}+\left(a_{d}+\sqrt{Z^{2}+b_{d}^{2}}\right)^{2}}}.
\end{align}
For the Galactic halo, \citet{irrgang_et_al:2013} proposed the following form:
\begin{align}\label{hallo}
        \Phi_{h}(R)=\begin{cases}
                 \frac{GM_{h}}{a_{h}}\left(\frac{1}{\gamma-1}\ln\left(\frac{1+\left(\frac{R}{a_{h}}\right)^{(\gamma-1)}}{1+\left(\frac{\Lambda}{a_{h}}\right)^{(\gamma-1)}}\right)-\frac{\left(\frac{\Lambda}{a_{h}}\right)^{(\gamma-1)}}{1+\left(\frac{\Lambda}{a_{h}}\right)^{(\gamma-1)}}\right), & \textrm{if \ensuremath{R<\Lambda},}\\
                -\frac{GM_{h}}{R}\frac{\left(\frac{\Lambda}{a_{h}}\right)^{(\gamma-1)}}{1+\left(\frac{\Lambda}{a_{h}}\right)^{(\gamma-1)}}, & \textrm{elsewhere,}
        \end{cases}
\end{align}
 which, after adopting $\gamma=2$, and choosing the first equation in (\ref{hallo})  because we do not allow a comet to go as far as $R=\Lambda=$200 kpc,  reduces to:

\begin{align}	
        \Phi_{h}(X,Y,Z) & =  \frac{GM_{h}}{a_{h}}\left(\ln\left(\frac{a_{h}+\sqrt{X^{2}+Y^{2}+Z^{2}}}{a_{h}+\Lambda}\right)-\frac{\Lambda}{a_{h}+\Lambda}\right).
\end{align}

In the absence of  other bodies, the equations of a comet and the Sun motion in the Galactic frame can be written as:
\begin{align}
\ddot{\vec{R}} &= - \nabla \Phi(\vec{R})  - \frac{G M_\odot}{r^3} \vec{r} \nonumber \\ &= \vec{F}_b(\vec{R})
+ \vec{F}_d(\vec{R}) + \vec{F}_h(\vec{R}) - \frac{G M_\odot}{r^3} \vec{r},  \\
\ddot{\vec{R}}_\odot &= - \nabla \Phi(\vec{R}_\odot) = \vec{F}_b(\vec{R}_\odot)
+ \vec{F}_d(\vec{R}_\odot) + \vec{F}_h(\vec{R}_\odot),
\end{align}
where $\Phi$ is the overall potential of the Galaxy  (\ref{eq:global_pot}),  and the subscripts of $\vec{F}$, similarly to those of $\Phi$, refer to their division into three commonly used parts: central bulge, disk, and halo.

The heliocentric motion of a comet results from
\begin{equation}\label{eq.helio_1}
\ddot{\vec{r}} = \ddot{\vec{R}} - \ddot{\vec{R}}_\odot =  - \frac{G M_\odot}{r^3} \vec{r} + \vec{f}(\vec{R}_\odot,\vec{r}),
\end{equation}
where
\begin{align}
\vec{f}(\vec{R}_\odot,\vec{r}) = \quad &\left[  \vec{F}_b(\vec{R}_\odot + \vec{r}) - \vec{F}_b(\vec{R}_\odot) \right] \nonumber \\
 +  &\left[ \vec{F}_d(\vec{R}_\odot+\vec{r}) - \vec{F}_d(\vec{R}_\odot)  \right] \nonumber \\
 +  &\left[ \vec{F}_h(\vec{R}_\odot+\vec{r}) - \vec{F}_h(\vec{R}_\odot)  \right]
=  \vec{f}_b + \vec{f}_d + \vec{f}_h.
\label{e13}
\end{align}

To avoid the subtraction of almost equal terms and the resulting loss of significant digits, the expressions for all parts of $\vec{f}$ need to be appropriately transformed.

Let us introduce the following quantities:
\begin{align}
p_1 &= \sqrt{R_\odot^2 + b_b^2}, \label{p1:def}\\
q_1 &= \sqrt{R^2 + b_b^2}, \label{q1:def} \\
p_2 & =  \sqrt{ X_\odot^2 + Y_\odot^2 + (a_d + p_3)^2}, \\
q_2 & =  \sqrt{ X^2 + Y^2 + (a_d + q_3)^2}, \\
p_3 & =  \sqrt{Z_\odot^2 + b_d^2}, \\
q_3 & =  \sqrt{Z^2 + b_d^2}, \\
\sigma & =  r^2 + 2 \, \vec{r} \cdot \vec{R}_\odot .
\end{align}
Then, the components of a comet's heliocentric acceleration per unit mass due to the Galactic tide can be expressed as:
\begin{align}
\vec{f}_b &=    \frac{G M_b}{q_1^3} \left[ \frac{\sigma\,\left( p_1^4 + q_1^4 + p_1^2 q_1^2\right)}{p_1^3 \,\left(p_1^3 + q_1^3 \right)} \, \vec{R}_\odot - \vec{r} \right], \\
\vec{f}_h & =   \frac{G M_h}{a_h R \left( a_h + R \right)} \left[ \left(1 + \frac{a_h}{R+R_\odot} \right) \frac{\sigma}{R_\odot \left( a_h + R_\odot \right)}\, \vec{R}_\odot
- \vec{r}\right], \\
\vec{f}_d & =    \frac{G M_d}{q_2^3} \left[  \frac{p_2^4 + q_2^4 + p_2^2 q_2^2}{p_2^3 \left( p_2^3 + q_2^3 \right)}
\left( \sigma  + 2 a_d \frac{z \left(z+ 2 Z_\odot\right)}{p_3+q_3} \right) \vec{R}_\odot  - \vec{r} + \vec{P}\right],
\end{align}
where
\begin{equation}\label{fpd}
\vec{P} =    \frac{a_d}{q_3} \left(
\begin{array}{c}
0 \\
0 \\
\frac{ A}{p_2^3 p_3 \left( p_2^3 p_3 + q_2^3 q_3\right)} Z_\odot - z \\
\end{array}
\right),
\end{equation}
and
\begin{equation}
A =  p_3^2 \left( \sigma + 2  a_d \frac{z \left(z+ 2 Z_\odot\right)}{p_3+q_3} \right) \left(p_2^4+ q_2^4 + p_2^2 q_2^2 \right)  + q_2^6 z \left(z+ 2 Z_\odot \right).
\end{equation}

The strict differential formulae can also be derived for an alternative model of the halo component of the Galactic potential, Model III in \citet{irrgang_et_al:2013}, which is deduced from the halo density profile of \cite[][hereafter NFW]{NFW:1997} , that is,
\begin{equation}\label{fibb}
\Phi_H(\vec{R}) = - \frac{G M_H}{R} \ln\left(1+\frac{R}{a_H}\right).
\end{equation}
For the NFW halo potential model, from Eq. (\ref{fibb}) we have
\begin{equation}\label{derfi}
\vec{F}_H = - \nabla \Phi_H(\vec{R}) = - \frac{G M_H}{R^3} \left[\ln\left(1+\frac{R}{a_H}\right) - \frac{R}{a_H+R} \right] \, \vec{R},
\end{equation}
and $\vec{f}_H = \vec{F}_H(\vec{R}) - \vec{F}_H(\vec{R}_\odot),$ should be transformed to
\begin{align}
\vec{f}_H &= \frac{G M_H}{R^3} \left[ W_1 \vec{R}_\odot + W_2 \vec{r} \right], \\
W_2 &= \frac{R}{a_H+R} - \ln\left(1+ \frac{R}{a_H} \right), \label{w2}\\
W_1 &= \ln\left( 1- \frac{\sigma}{\left(a_H+R\right)\left(R_\odot+R\right)}\right) - \frac{a_H \sigma  R}{\left(a_H+R\right)\left(a_H+R_\odot\right) R_\odot^2} \nonumber \\
&  + \frac{\sigma \left(R_\odot^2 + R^2 + R_\odot R\right)}{\left(R_\odot+R\right) R_\odot^3} \left[ \ln\left(1+\frac{R_\odot}{a_H}\right) - \frac{R R_\odot}{\left(a_H+R\right)\left(a_H+R_\odot\right)}\right].~~ \label{w1}
\end{align}
If it does not degrade the accuracy (e.g. in quadruple arithmetic), we can use  a simpler form of the first logarithm  in $W_1$:
\begin{equation}\label{alt}
\ln\left( 1- \frac{\sigma}{\left(a_H+R\right)\left(R_\odot+R\right)}\right) = \ln\left(\frac{a_H+R_\odot}{a_H+R}\right).
\end{equation}
The derivation of the above equations is elementary. The outline can be found in Appendix~\ref{Ap:der}.

If one aim is to study only the Galactic tide influence on Solar System bodies, the above proposed formulae offer a fast and accurate solution. However, they require galactocentric positions of the Sun $\vec{R}_{\odot}(t)$. It is possible to construct more or less approximate analytical formulae to achieve this, for example, by obtaining a polynomial approximation of the results of direct numerical integration of the Sun's galactocentric motion. Our tests showed that such ephemeris can be quite compact. For example, the backward motion of the Sun in the Galaxy spanning 30 Myr can be successfully described by six subsequent intervals covered by approximating polynomials of  degree 5, individually for each coordinate. In this case. the whole backward ephemeris of the Sun takes only 120 double precision numbers. We offer an example of such an ephemeris in Appendix~\ref{app-ephem}.

Since we are interested in both Galactic and stellar perturbations, we decided to prepare a compound source for the Sun and star positions, described in Sect.~\ref{ephemeris}.

\section{Simultaneous Galactic and stellar perturbations\label{sect:galstel}}

When searching for the source region or regions of the observed LPCs, we have to study their past dynamics, starting from their "original"\ orbit and taking into account both Galactic and stellar perturbations. There is an ongoing project lead by Królikowska and Dybczyński, aimed at revealing the dynamical history of the observed LPCs; for more, see \cite{kroli-dyb:2020} and references therein.

Adding stellar perturbations to the heliocentric comet equations of motion is much simpler than carrying out this process for Galactic tides. It is reduced to including a sum of point-mass interactions with all stars described by their heliocentric positions. This requires an extension of (\ref{eq.helio_1}) into
\begin{equation}\label{eq.helio_2}
\ddot{\vec{r}} =  - \frac{G M_\odot}{r^3} \vec{r} + \vec{f}(\vec{R}_\odot,\vec{r}) + \sum_{s=1}^{N}{\left[\frac{GM_s(\vec{r_s}-\vec{r})}{|\vec{r_s}-\vec{r}|^3}\right]},
\end{equation}
where $\vec{r}_s=\vec{R}_s - \vec{R}_\odot$, and the summation in the last term is over all $N$ stars included in the model. The use of this equation requires the knowledge of the positions of the Sun and of all the stars in the galactocentric frame at each moment covered by the numerical integration.
Obtaining these Sun and star positions requires a numerical integration of their galactocentric motion that takes into account their mutual interactions and an overall Galactic potential. We note that this should not be repeated each time we investigate a subsequent comet. For this purpose, we decided to build a numerical ephemeris of the Sun and stars.

\subsection{Numerical ephemeris of the Sun and the stars galactocentric motion}\label{ephemeris}

For the numerical integration, we use the well-known and widely recommended Gauss-Radau integrator published  by \cite{everhart-ra15:1985}.
Although we have tested various orders of the integrator, the practice shows that the classical 15th-order RA15 works best
for the numerical integration of a comet motion in the heliocentric frame.
The integrator implements a collocation method, which means that at every step (referred to as a "sequence" by Everhart), a polynomial
of time is constructed to approximate the integrated body motion. Although the polynomial is evaluated only at the end of each sequence,
its coefficients may also serve to create a dense output: similarly to the Taylor series methods, the collocation methods are capable of providing the position and velocity at any time with no additional cost (even though the accuracy between the endpoints is formally lower than in the Taylor polynomial case).
We decided to use this property of the Gauss-Radau integrator to construct the numerical ephemeris by simply recording all polynomial coefficients at the end of each sequence when integrating the Sun and stars in the galactocentric frame.

The list of potential stellar perturbers used to construct an ephemeris is based on the publicly available stellar database\footnote{\url{https://pad2.astro.amu.edu.pl/StePPeD}} (version 2.3) announced by \cite{rita-pad-magda:2020}. To make our proposition more up to date, we refined this list using the latest {\it Gaia} EDR3 catalog \citep{GaiaEDR3-summary:2021} and making this partial update publicly available as the StePPeD release 3.0.  It is worth mentioning that thanks to the more accurate and complete data in the {\it Gaia} EDR3 catalog, we identified 43 additional double stars. As a result, the number of double or multiple systems included in our list of potential perturbers increased to 81. In all available cases, the center of mass of the system is used. The work on the complete update of the whole StePPeD database by incorporating {\it Gaia} EDR3 results which also include some new stars is in progress.  As a result of this refinement, the whole number of stars or stellar systems taken into account is 407,  but in order to optimize the ephemeris file size and the speed of its use, we decided to limit the number of perturbers by the following rules.

First of all, we produce two separate ephemerides: one for integrating a comet motion backward and one for the forward integration. This is fully justified because studying cometary past dynamics (starting from the "original"\ orbit) and its future dynamics (starting from the  "future"\ orbit) are always separate tasks.

Moreover, most of the perturbers are now much farther than the 4~pc limit assumed when constructing the database. As a result,  in the ephemeris dedicated for forward integrations of the Sun, we include 25 stars or stellar system that are currently closer than 4~pc, and 175  perturbers that will pass the Sun closer than 4~pc in the future, which makes the total number of perturbers in this ephemeris equal 200 plus the Sun. In the same manner, but looking to the past, we restricted the number of perturbers in the backward ephemeris to  $25+207=232$ objects plus the Sun.

\begin{figure}
  \includegraphics[scale=0.75]{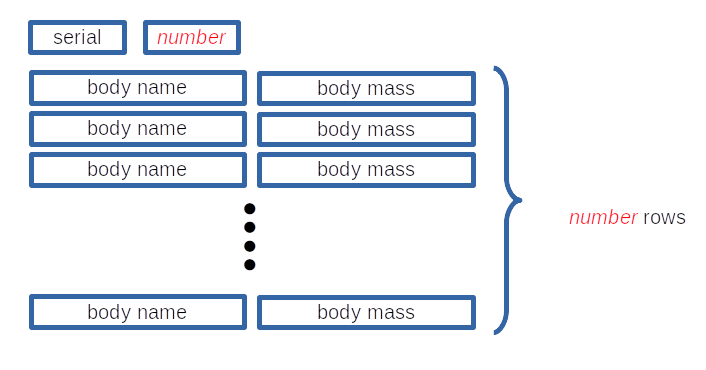}

  \caption{Structure of the header part of the binary ephemeris file; "Serial" is the (4 byte integer) code of the ephemeris version (1255 at  present), "number" is the (4 byte integer) number of bodies included. They are followed with the table of  "number" rows, containing 8 byte character string with the perturber name and double precision floating point number containing the mass of the perturber.\label{fig-header}}
\end{figure}

For the purpose of storing ephemeris data in binary files, we used integer numbers (4 bytes), double precision floating point numbers (8 bytes) and 8 byte character strings for perturber names. Both ephemerides have the same format of the binary file. It starts with the header: two integer numbers and a table of perturber names and masses, structured as described in Fig.\ref{fig-header}.

\begin{figure}
        \includegraphics[scale=0.75]{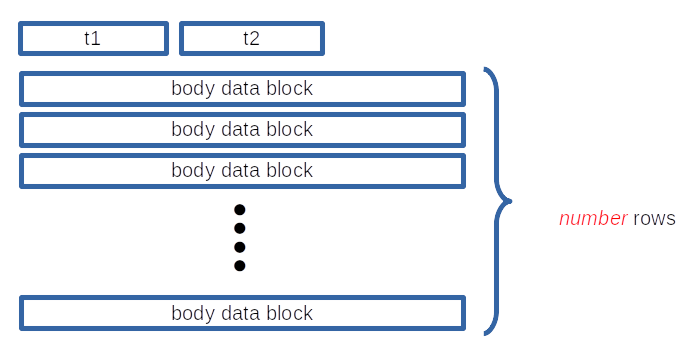}
        \caption{Structure of the "one-step data blocks" of the binary ephemeris file; t1 and t2 are the ends of the validity time interval of polynomial coefficients stored in "body data blocks{\it }." The number of these blocks equals the number of bodies included in the ephemeris. Each "body data block" is 240 bytes long (3x10 double precision floating point numbers).\label{fig-data}}
\end{figure}

Next, the main body of the ephemeris follows. It consists of a large number of "one-step data blocks{\it }." Each of them contains two epochs, expressed as Julian Days and stored as double precision floating point numbers. These are the ends of the validity time interval of polynomial coefficients stored in the rest of one-step data block{\it }, see Fig.\ref{fig-data}.

Polynomial coefficients are stored in body data blocks{\it }, each containing 10 double precision floating point numbers for each of the three coordinates of the body in the galactocentric frame, expressed in parsecs. As a result, each body data block is 240 bytes long. In the backward ephemeris there are 233 body data blocks in each one-step data block{\it }. In the forward ephemeris the number of bodies is only 201.

We decided for both ephemerides to cover 30~Myr, which is comparable with the orbital period of a comet having the semi-major axis of 100\,000~au, but only for a handful of observed LPCs we have to perform a numerical integration spanning more than 10 Myr. As a result we have 2323 one-step data blocks
  in the backward ephemeris, and 2382 in the forward one. The total size of the ephemerides files is 128 MB and 115 MB, respectively. The average step size is on the level of 10\,000 years but it goes down as low as to ten years during the closest encounters.

It should be noted here that these ephemerides were obtained from the precise numerical integration (80 bits arithmetic, i.e., long double in the C language) of the stars and the Sun in the galactocentric frame under the overall Galactic potential and taking into account all mutual gravitational interactions. The internal precision was kept on the level of $10^{-11}$ in terms of the galactocentric positions in parsecs, which gives 15 significant digits. Starting data for all bodies are based on nominal stellar data taken in most cases from the {\it Gaia} EDR3 catalog  and are referred to the epoch 2016.0. For such details as the Sun initial conditions, numerical parameters, frame definition, or dynamical model, we refer to \cite{dyb-berski:2015}.

The stellar ephemeris proposed here is designed for screening tests when searching for detectable star-comet interactions, using long lists of comets and potential stellar perturbers. It can also be used when analyzing the influence of the cometary data uncertainties. However, finally, when investigating a certain case of the close encounter, it is necessary to also take into account the stellar data uncertainty that cannot be performed with this ephemeris, based on nominal stellar data.

Both ephemerides binary files as well as the documentation and examples of a code using them are publicly available on the Internet\footnote{\url{https://pad2.astro.amu.edu.pl/StePPeD}} as a part of our stellar database site (see the download section). The ephemerides (and the code if necessary) will be updated each time we update our database of potential stellar perturbers. 

\begin{table*}
        \caption{Five stellar perturbers with the strongest impact on the past motion of all analyzed LPCs.\label{tab-stars}}
        \begin{tabular}{c c r r r c c r r}
                \hline
                StePPeD & common & \multicolumn{1}{c}{{\it Gaia} EDR3} & \multicolumn{2}{c}{change in $q_{\rm prev}$} & \multicolumn{2}{c}{solar proximity parameters} & star mass &  rel. vel.\\ 
                ID & name & \multicolumn{1}{c}{ID} & >10\% & >50\% & mindist [pc] & mintime [Myr] & [M$_{\odot}$] &  km s$^{-1}$\\
                \hline
                &&&&&&&&\\              
                P0230 & HD~7977        &  510911618569239040  & 145 & 134 & $0.027\pm0.065$ & $-2.80\pm0.04$ & 1.08 &  $27.1\pm0.4$\\
                P0509 &                &   52952720512121856  &  63 &   8 & $0.256\pm0.183$ & $-0.55\pm0.07$ & 1.46 & $38.7\pm3.7$\\
                P0506 &                & 5571232118090082816  &  37 &   4 & $0.199\pm0.012$ & $-1.08\pm0.01$ & 0.77 & $90.2\pm0.3$\\
        P0417 & Ton 214        & 1281410781322153216  &  10 &   0 & $0.514\pm0.025$ & $-1.47\pm0.01$ & 0.85 & $32.7\pm0.1$\\
                P0508 &                & 2946037094762449664  &   8 &   3 & $0.240\pm0.508$ & $-0.92\pm0.11$ & 0.25 & $43.0\pm3.8$\\
                \hline
        \end{tabular}
\end{table*}

\subsection{Computer implementation for cometary dynamics}

As described in \cite{kroli-dyb:2020} for each studied LPCs we would like to follow their motion to the past, starting from their "original"\ orbits, and to the future, starting from "future"\ orbits. These integrations are typically stopped at the previous or next perihelion, or when a comet leaves a heliocentric sphere of the 120\,000 au radius. The previous perihelion distance value can be used to discriminate between dynamically old and new comets.

While preparing the paper cited above, we encountered problems with these calculations resulting from losing significant digits due to a very small difference between the Sun and comet positions expressed in the galactocentric frame. This enforced us to perform a high accuracy numerical integration which appeared to be extremely time consuming. As a result, we decided to switch off all mutual interactions between stellar perturbers in the dynamical model. This greatly speeded up our calculations and in the great majority of cases the resulting comet "previous"\ and "future"\ orbits are correct. But there are several cases where the mutual inter-stellar gravitational action changes a particular perturber trajectory in a manner, which  changes the resulting "previous"\ or "next" orbits slightly. Those problems stimulated us to develop the alternative methods described in this paper.

During the ephemeris production, when integrating backwards, we recorded nine encounters between stars closer than 0.2~pc, and only two of them are the encounters with the Sun.  Three of the encounters were closer than 0.1~pc: stars \object{{\it Gaia} EDR3 2946037094762449664} and \object{{\it Gaia} EDR3 5261593808165974784}  
approached each other at a distance of 0.05~pc 0.89 Myr ago, the stars \object{TYC~160-136-1} and  \object{{\it Gaia} EDR3 5949463366461232896} miss distance was 0.09~pc 1.7 Myr ago, and the star HIP~31626 passed near the double star \object{{\it p Eridani}} at a distance of 0.04~pc 0.42 Myr ago. It is worth noting that the same star  \object{HIP~31626} met another one, \object{{\it Gaia}~EDR3 2730508416002618752} a little bit later. When analyzing the future motion we recorded seven mutual star encounters below 0.2~pc. In the rare cases of such close encounters, even small modifications of stellar trajectories (on the level of 0.1~pc) can induce substantial changes in cometary motion, as shown in \cite{First-stars:2020}.  It should be stressed that the above data for the closest encounters are obtained from the nominal stellar data, which have a different level of  accuracy.

With the exact differential formulae described in Sect.~\ref{differential}, and having numerical ephemerides containing the Sun and the potential stellar perturbers' positional galactocentric data, the construction of a computer program for studying the comet's long-term motion is not difficult. Preparing the code for Galactic tides is quite straightforward, save for one exception: if we uses the NFW variant of the Galactic halo model, it is necessary  to use the special numerical function for calculating $\ln(1+\xi)$ when the value of $\xi$ is small (e.g., function \ \texttt{log1p} in the C/C++  language mathematical library) to evaluate Eqs. (\ref{w1}) and (\ref{w2}).

It is also worth observing that the step size of the stellar ephemeris is typically several orders larger than the step size in the comet heliocentric motion integration. For the optimal use of the ephemeris, the calling function should use static variables to monitor whether a new set of coefficients should be read from a file.

In our code, we use internally dedicated units for keeping the optimal numerical precision, namely: a solar mass, $10^3$ years, and $10^4$ au. When reading the ephemeris file, the rescaling is necessary as well as the transformation of stellar positions from galactocentric to heliocentric, with the Sun position included.

In all the above, we use the term "heliocentric"\textit{} for simplicity but, in fact, we work within the frame, with the origin at the barycenter of the Solar System. It is a widely accepted practice that the "original"\ and "future"\ comet orbits are presented in a barycentric frame (see \cite{kroli-dyb:2020} for details) and we keep our calculations of the "previous"\ and "next"\ orbits in the same reference frame. As a result, we used a central body mass equal to 1.001341838222754 (value taken from the DE430 JPL planetary ephemeris constants, see \cite{folkner_et_al:2014}).

The comparison with our previous, direct galactocentric integrations of a comet motion showed the improvement of an internal precision by at least six orders of magnitude.  Moreover, the present code works over 100 times faster. It is worth mentioning that obtaining in the galactocentric integration the accuracy comparable to the present code requires quadruple precision, which is approximately 200 times slower.

\section{Overall stellar impact on the cometary orbit evolution}

Using the proposed methods we performed an overall test of the stellar perturbations effect on the past motion of LPCs. For all 270 preferred original comet orbits taken from the version 3.0 of the CODE catalog \citep{kroli-dyb:2020}, we performed a series of numerical integrations. Each comet was first integrated under the influence of the Galactic tide and all 232 stellar perturbers taken from the release 3.0 of the StePPeD database. Then we repeated this calculation 232 times with each individual stellar perturber removed. We compared the comet previous perihelion distance obtained from the full model and the one with a particular star omitted. In this way, we reasonably measured the influence of each individual perturber on each comet's past motion.

In Table~\ref{tab-stars}, we list five stellar perturbers which showed the strongest influence on the analyzed sample of LPC orbits. The first three columns contain the star identifications. In the fourth column, we present the number of comets for which we observed a change in the previous perihelion distance greater than 10\% in the numerical experiment described above. The same, but for changes greater than 50\%, is showed in the fifth column. Next three columns show the parameters of each particular star closest passage near the Sun, and the estimate of its mass used in our calculations. In the last column we show the relative velocity at the moment of the star closest approach.

Solar proximity parameters together with their uncertainties were obtained from a series of the extensive numerical Monte Carlo simulations. In the precise numerical integration performed in the galactocentric frame, we use all 232 perturbers and take into account all their mutual interactions. For the uncertainty estimation we repeated this calculation replacing a particular star with at least of 10\,000 of its clones (using nominal data for the remaining objects). These clones were drawn according the the six-dimensional covariance matrix from {\it Gaia} EDR3\footnote{see the {\it Gaia} EDR3 documentation, paragraph 4.1.7.0.4 on page 196}. For each perturber we generated more than 10000 of its clones, numerically integrated their past motion, and stopped them at their closest approach to the Sun.  This method of calculating the uncertainty estimation is obviously simplified since we take into account only one star's uncertainties, while the rest move along their nominal tracks. Fortunately, the close star-star encounters are very rare (and frequently with a high relative velocity), so this simplification still offers the correct order of the scatter of the results.  As it is described in detail in  \cite{dyb-berski:2015}, we calculated the coordinates of the clone cloud centroid (calculating a simple mean in each heliocentric coordinate, i.e., $\bar{x}$,  $\bar{y}$,  $\bar{z}$), and then obtaining its distance from the Sun:
\[mindist=\sqrt{\bar{x}^2+\bar{y}^2+\bar{y}^2}.\]
As the uncertainty of this expected minimal distance, we measure the radius of a sphere around the centroid ($\bar{x}$,$\bar{y}$,$\bar{z}$) containing 90\% of individual clone proximity points. The epochs of the closest approach are presented in Table~\ref{tab-stars} as their mean and standard deviation since the distribution of these results is close to Gaussian in all five cases.

\begin{figure}
        \includegraphics[angle=270,width=1\columnwidth]{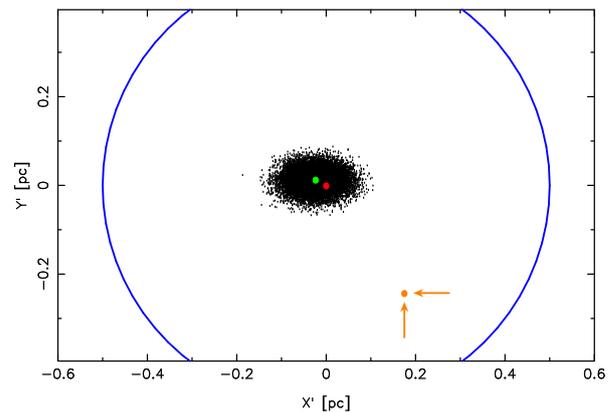}
        \caption{Distribution of the cloud of 16\,545 clones of P0230 stopped at the closest proximity to the Sun about 2.80 Myr ago. All points are projected onto the maximum scatter plane X'Y'. The green dot marks the nominal approach position, while the red one depicts the position of the Sun. The blue circle shows the widely adopted approximate boundary of the Oort cometary cloud at 0.5~pc from the Sun. Orange dot and arrows show the position of C/2015~XY1 at the moment of the nominal P0230 approach; see Sect. \ref{sect-example} for details. \label{P0230_scatter}  }
\end{figure}

If the presented uncertainty in the minimal distance is greater than the value of the expected proximity distance (as in the two cases included in Table~\ref{tab-stars}), we should interpret this fact by expecting some fraction of clones to pass on the other side of the Sun than the centroid. In other words, the position of the Sun is inside the sphere of 90\% of clones, and as a result, an arbitrarily small proximity distance is statistically possible. We show an example of such a situation in a case of the first and the most important star from Table~\ref{tab-stars}, namely, P0230 (\object{HD 7977}); see Fig. \ref{P0230_scatter}.

The possibility of a close approach of this star to the Sun  was first pointed out by \cite{Bailer-Jones:2018}. In earlier papers, this star was omitted because its parallax was first measured by {\it Gaia} mission and published in the {\it Gaia} DR1 catalog \cite{GaiaDR1:2016} and its radial velocity appeared only in {\it Gaia} DR2 \citep{Gaia-DR2:2018}. In \cite{Bailer-Jones:2018}, based on the {\it Gaia} DR2 data, the authors estimated the nominal closest distance of P0230 to the Sun as 0.43~pc.

That trajectory of P0230 was recognized by \cite{First-stars:2020} as passing very close to the long period comet C/2002~A3 LINEAR, possibly imparting a strong gravitational influence on the comet motion.
Following updates to the {\it Gaia} EDR3 data \citep{GaiaEDR3-summary:2021}, the situation has changed substantially. Now the nominal minimal Sun-star distance is only 0.03~pc (see the first row in Table \ref{tab-stars}), and based on Fig. \ref{P0230_scatter}, we clearly see that due to the still significant uncertainties in the stellar data for P0230, this distance might happen to be arbitrarily small or significantly larger, and the geometry of this passage is highly uncertain.

According to the data in Table~\ref{tab-stars}, 145 of LPCs listed in the CODE \citep{kroli-dyb:2020} database might have had their previous perihelion distance changed by more than 10\% due to the gravitational action of this star, and 134 of them by more than 50\%. Taking into account that only 129 comets from our list have their previous orbital period longer than 2.8 Myr, it is evident that all comets that exhibited a chance of an encounter with P0230 via our calculations were affected.

The importance of this perturber comes from its relatively large mass, estimated to be equal to 1.08 M$_{\odot}$ \citep{TIC-8:2019}, but mainly from its very close flyby near the Sun. This is a well-known effect that the change of a comet orbit is the result of the difference between a gravitational pull induced by a passing star on a comet, and that gained by the Sun. In the case of such a close P0230 flyby, effectively, all comets can be indirectly affected. This important stellar passage near the Sun and its possible consequences will be studied in detail in a separate paper which is in preparation.

In the next section, we present a detailed analysis of an individual example: the past dynamical evolution of the comet C/2015~XY1 Lemmon. It properly illustrates the cumulative influence of two different stars listed in Table~\ref{tab-stars}.

\begin{figure}
        \includegraphics[angle=270,width=1\columnwidth]{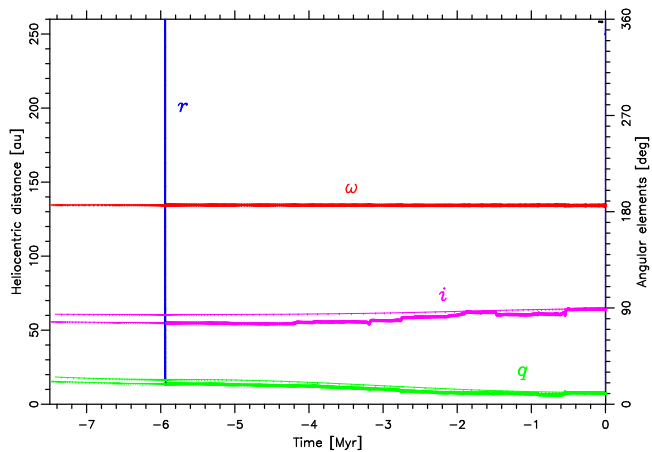}
        \caption{ Dynamical past evolution of the nominal orbit of C/2015~XY1. Here P0506 and P0230 are excluded from the model. Changes in a perihelion distance (green), an inclination (fuchsia) and an argument of perihelion (red) are shown. The thick lines depict the result of the full dynamical model while thin lines show the evolution of elements in the absence of any stellar perturbations. Angular elements are expressed in the Galactic frame.\label{2015xya5_bez_obu}}
\end{figure}
\begin{figure}
        \includegraphics[angle=270,width=1\columnwidth]{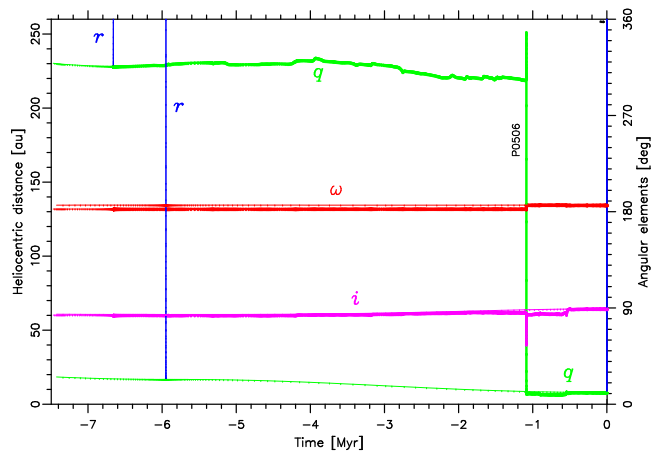}
        \caption{ Past evolution of of C/2015~XY1 with only the star P0230 excluded. The strong interaction with P0506 is evident.\label{2015xya5_bez_230}}
\end{figure}
\begin{figure}
        \includegraphics[angle=270,width=1\columnwidth]{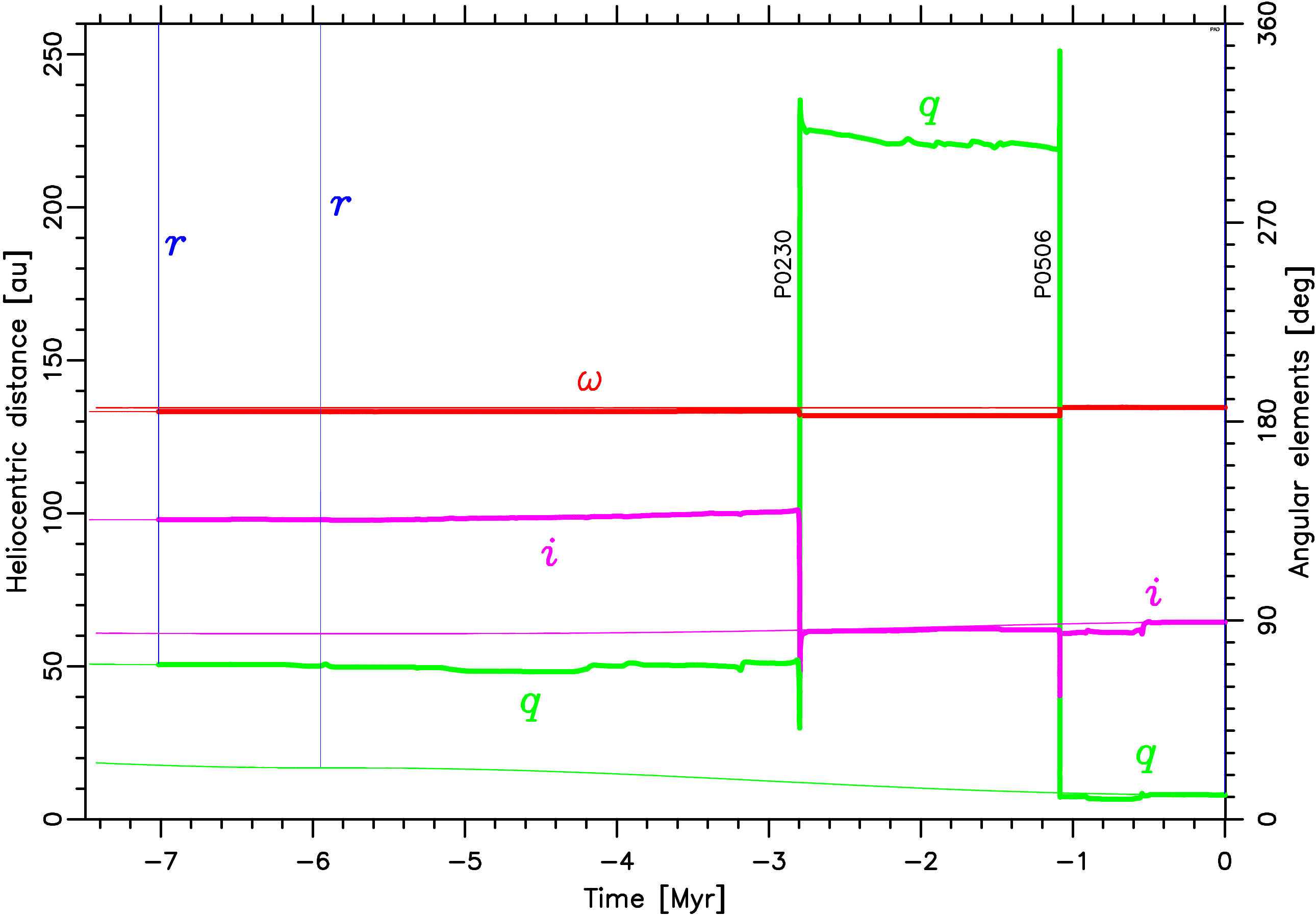}
        \caption{  Past changes in nominal orbital elements of C/2015~XY1
with all potential stellar perturbers included. The strong interaction of C/2015~XY1 with P0506 and P0230 is shown.\label{2015xya5_all}}
\end{figure}
\begin{figure}
        \includegraphics[angle=270,width=1\columnwidth]{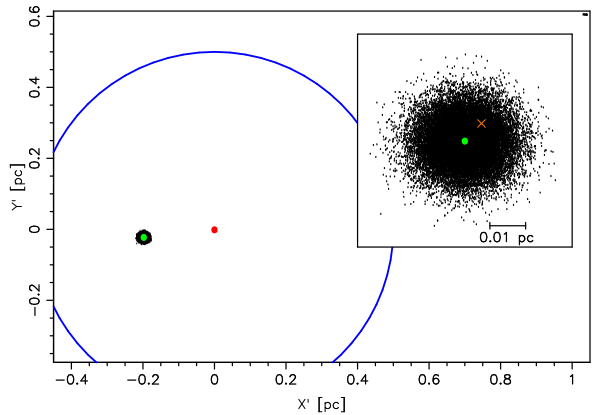}
        \caption{Distribution of the cloud of 25\,000 clones of P0506 stopped at the closest proximity to the Sun 1.09 Myr ago. All points are projected onto the maximum scatter plane X'Y'. The green dot marks the nominal approach position, while the red one depicts the position of the Sun. The blue circle shows the widely adopted approximate boundary of the Oort cometary cloud at 0.5~pc from the Sun. An inset at the upper right shows the star clones cloud enlarged ten times to show the position of C/2015 XY1 (an orange cross) at the closest approach. \label{P0506_scatter}  }
\end{figure}

\section{Example result in detail: past dynamics of comet C/2015~XY1 Lemmon}\label{sect-example}

When reviewing the effect of stellar perturbations on all LPCs included in the CODE database, we found a very interesting case of the past motion of C/2015~XY1 Lemmon. This comet has an orbit of the greatest quality 1a+ (see \cite{Kroli-Dyb:2018b} for an explanation). As the starting point we used the preferred original orbit of this comet taken directly from the CODE database \citep{kroli-dyb:2020} For convenience, we quote the elements of this orbit in Table \ref{tab-elements}. We found that during the orbital period preceding the observed perihelion passage, this comet significantly changed its orbit  twice, as a result of perturbations caused by stars P0506 and P0230. Based on nominal data, it was 1.09 Myr ago that P0506 passed 0.005~pc ($\sim$1100~au) from this comet. However, much earlier, that is, 2.8 Myr ago, when the comet was near its aphelion, it received another change of its heliocentric orbit as a result of the strong interaction between the Sun and P0230.

\begin{table}
        \caption{Original orbital elements of the preferred solution (a5) for C/2015~XY1, taken from the CODE database. The angular elements are referred to the ecliptic of J2000.\label{tab-elements}}
        \begin{tabular}{l l}
                \hline
                Epoch & 1707 09 30      
\\
                Perihelion date &       2018 04 29.1664$\pm$0.0019
\\
                Perihelion distance [au] &      7.935467$\pm$0.000009
\\
                Eccentricity &  0.999757$\pm$0.000006
\\
                Argument of perihelion [$^{\circ}$] &   196.23113$\pm$0.00012
\\
                Ascending node [$^{\circ}$] &   281.55378$\pm$0.00001
\\
                Inclination [$^{\circ}$] &      148.84193$\pm$0.00002
\\
                Recip. semi-major axis [10$^{-6}$ au$^{-1}$] &  30.57$\pm$0.67 \\
                \hline
        \end{tabular}
\end{table}

If we exclude  these two events from our simulation, the past dynamical evolution of C/2015~XY1 would look like in Fig.\ref{2015xya5_bez_obu}. All perturbations from the other 230 stars were taken into account here, but they appeared to be small. In Figs \ref{2015xya5_bez_obu}, \ref{2015xya5_bez_230}, and \ref{2015xya5_all}, we keep the same scales for a simpler comparison. The left vertical axis is expressed in au and corresponds to the osculating perihelion distance plot ($q$, green line) as well as the heliocentric distance plot ($r$, thin, vertical blue line, visible only at the previous perihelion). The right vertical axis is expressed in degrees and describes the evolution of the osculating inclination ($i$, fuchsia line) and of the argument of perihelion ($\omega$, red line). Both these angular elements are expressed in the Galactic reference frame. Thick lines show the dynamical evolution under the simultaneous stellar and Galactic actions, while thin lines depict the evolution when all stellar perturbations are excluded. In Fig. \ref{2015xya5_bez_obu}, it is clearly shown that after excluding the two most important perturbers, P0506 and P0230, the comet orbital elements evolve very similarly to the case when only  Galactic tide is considered.

However, as it was mentioned above, our calculations show that 1.09 Myr ago this comet motion was strongly perturbed by the star P0506. The data in the {\it Gaia} EDR3 catalog for this star are much more precise than before, so its trajectory near the Sun is much better defined, as it is shown in Fig.\ref{P0506_scatter}. Additionally, a new and more precise radial velocity of this star was obtained by \cite{Errmann:2020}. In Fig.\ref{P0506_scatter}, we show 25\,000 clones of this star (black dots) stopped at their closest approach to the Sun, and projected onto the $X'Y'$ maximum scatter plane obtained by a principal component analysis (see \cite{dyb-berski:2015} for details). A green dot marks the nominal position of the star at its closest approach, a red dot shows the position of the Sun. An inset at the right upper corner shows the cloud of the star clones enlarged ten times and the position of C/2015~XY1 (the orange cross) at the moment of the closest star-comet encounter, projected onto the same $X'Y'$ plane. The uncertainty of a comet position is below 0.001~pc ( $\sim$200~au). This uncertainty  was obtained by the numerical integration of 5000 clones (hereafter virtual comets, VCs) of C/2015~XY1. The VCs were generated by the method proposed by \cite{sitarski:1998}, based on the osculation orbit of C/2015~XY1 and then propagated back in time to obtain a set of original orbits reflecting the observational uncertainties in comet positional observations. The set of these VCs can be downloaded from the CODE catalog web page\footnote{\url{https://pad2.astro.amu.edu.pl/CODE}}. Such an uncertainty estimation can be done quite simply and fast, using the stellar ephemeris described in Sect.~\ref{ephemeris}.

The comet in Fig.~\ref{P0506_scatter} is slightly above the $X'Y'$ maximum scatter plane and its nominal minimal distance from this star is only about 1120~au. 90\% of star clones are situated in the sphere with a radius of 0.013~pc ($\sim$2000~au) around the P0506 nominal proximity point. The cometary and stellar uncertainties taken into account simultaneously result in the minimal comet-star distance in the range of 13 -- 5840~au obtained from a set of 12\,000 random pairs of star and comet clones.

The nominal orbit evolution with perturbations from P0506 taken into account (i.e., now only P0230 is excluded) is shown in Fig.\ref{2015xya5_bez_230}. One can notice a strong star-comet interaction 1.09 Myr ago. The angular elements do not change significantly, but an orbital period increases to almost 7 Myr, and the previous perihelion distance increases from 14~au to $\sim$230~au. This last effect might have been interpreted as a proof that this comet is dynamically old, but this is not true, since another significant orbit change occurred about 2.8 Myr ago due to P0230.

This nominal orbit evolution,  compatible with our current knowledge,  is depicted in Fig.\ref{2015xya5_all}. There we see the significant C/2015~XY1 orbit change 2.8 Myr ago. The orbital period is again a little longer, there is a big change in the inclination, and the perihelion distance is significantly cut down to $\sim$50~au. However, the orbital evolution, depicted in Fig. \ref{2015xya5_all}, should be interpreted only as one of the possible scenarios due to the still great uncertainties on the P0230 passage geometry near the Sun.

\section{Influence of the stellar and cometary data uncertainties}

\label{sec:cloning}
\begin{figure}
        \includegraphics[angle=270,width=1\columnwidth]{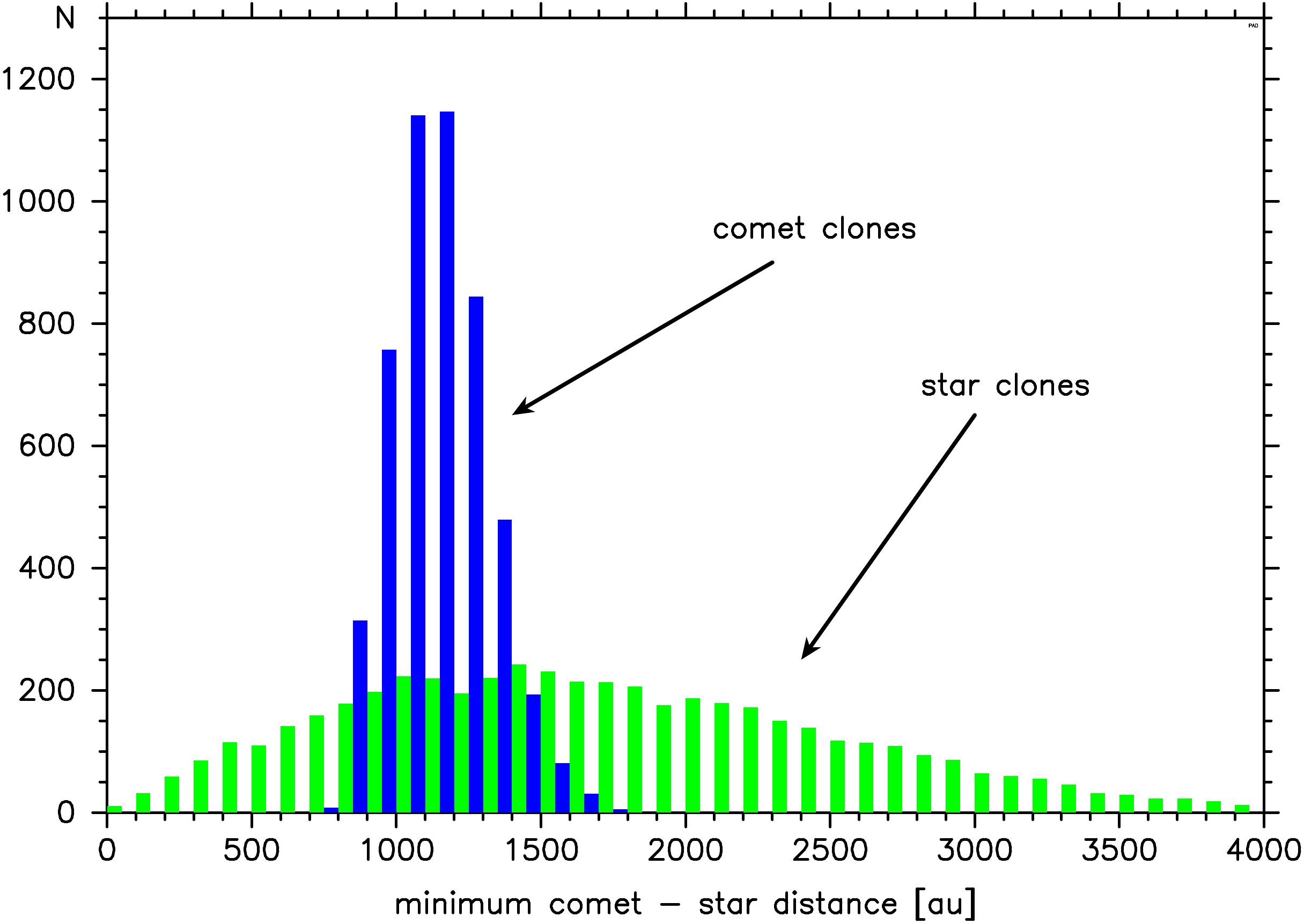}
        \caption{Comparison of the cometary and stellar data uncertainty influence on the minimum star-comet distance distribution.\label{fig_min_dist_uncert}}
\end{figure}

\begin{figure}
        \includegraphics[angle=270,width=1\columnwidth]{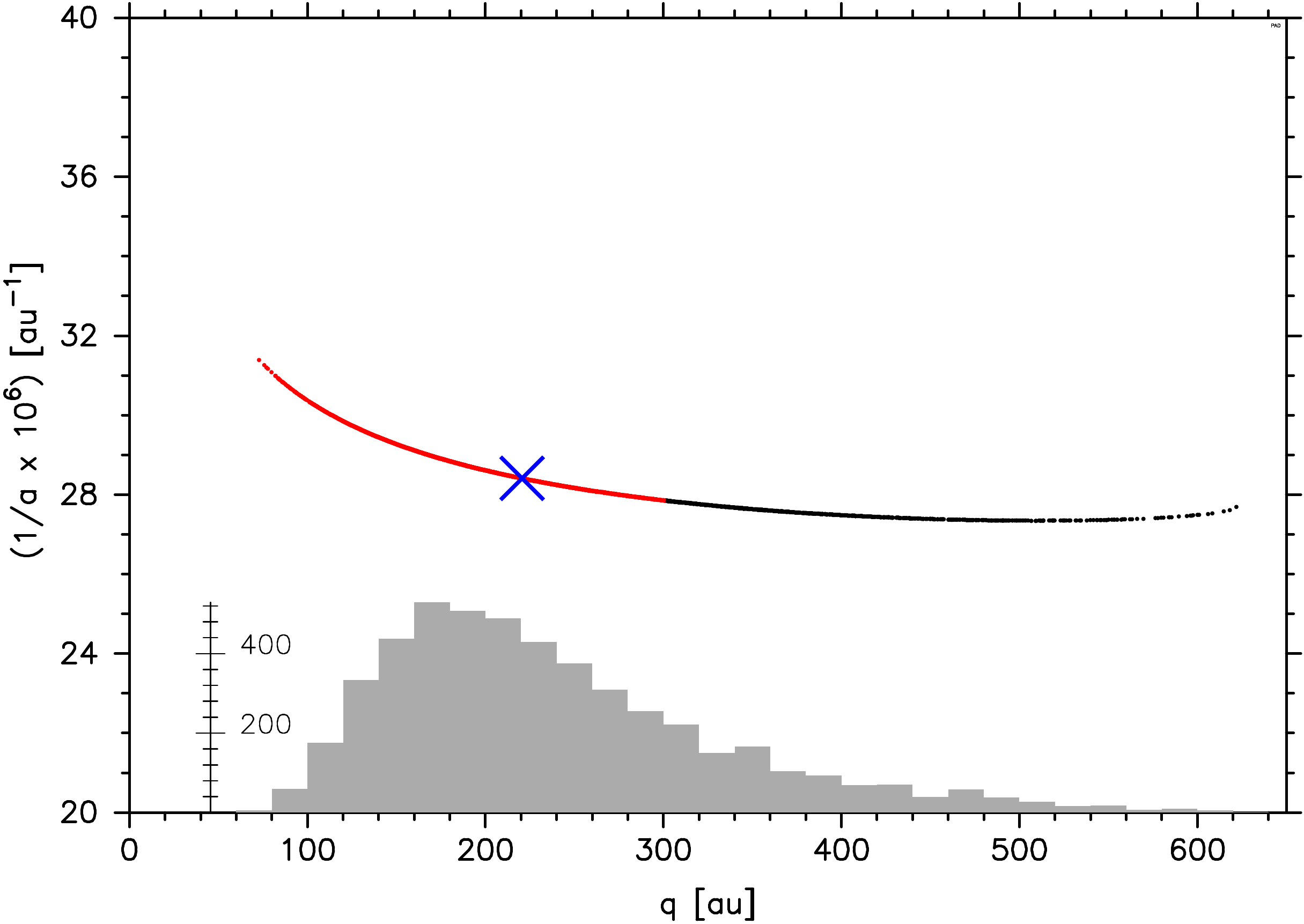}
        \caption{Effect of the cometary data uncertainty on C/2015~XY1 orbital elements recorded 2.0 Myr ago. See text for a detailed explanation.\label{fig_comet_uncert}}
\end{figure}

\begin{figure}
        \includegraphics[angle=270,width=1\columnwidth]{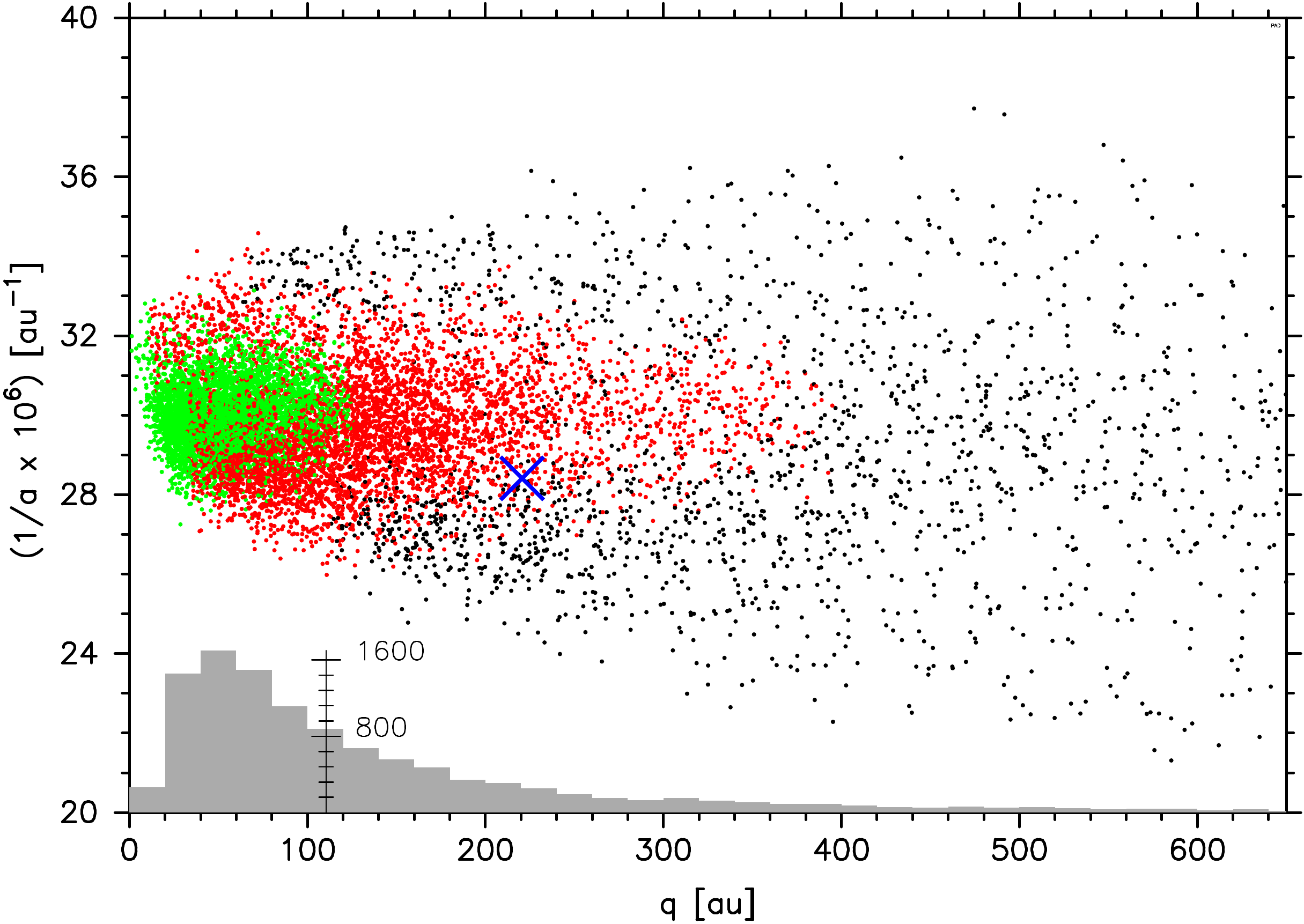}
        \caption{ Simultaneous effect of both cometary and stellar data uncertainties. The scales are the same as in Fig. \ref{fig_comet_uncert} \label{fig_combined_uncert}}
\end{figure}

Taking into account the relatively large uncertainties of stellar data for P0230, let us focus on the influence of the data uncertainty in the case of the close encounter between  C/2015~XY1 and P0506. This encounter occurred 1.09 Myr ago, so we decided to analyze osculating elements of this comet orbit exactly at 2.0 Myr ago and to show the individual and combined effect of the uncertainties.

Before showing the distributions of orbital elements, we present the comparison between the influence of only cometary and only stellar data uncertainties on the minimum  star-comet distance in the studied case. In Fig. \ref{fig_min_dist_uncert}, we show two distributions of the minimum star-comet distances. The blue one is based on 5000 VCs of C/2015~XY1 obtained as described in the previous section and the nominal stellar data. The green one shows the distribution of the same parameter, but based on random 12\,000 star clones and the comet nominal data (before plotting, the histograms were normalized to the same vertical scale). The blue histogram extends from $q=792$ to 1797~au with the maximum around the nominal value of 1100~au. The green histogram extends from $q=12$ to 5573~au, and 146 clones from the right tail are out of the border of the plot. The maximum value for the green distribution is significantly larger and lays somewhere between 1200 and 1600~au. It is worth mentioning that the distribution of the minimum star-comet distance, when both a comet and a star were randomly drawn from their respective sources, is almost identical to the green one, and extends from $q=13$ to 5837~au (with 152 clones above 4000~au). We do not present this distribution, since both histograms would overlap.

To study the spread of osculating elements at 2.0 Myr ago, we also started by analyzing the effect of the cometary data uncertainty. In Fig. \ref{fig_comet_uncert}, we present the plot of osculating $1/a$ versus $q$ for all 5000 VCs of C/2015~XY1 augmented with the marginal distribution of the perihelion distance, which is the most perturbed element during the P0506 passage (see Fig. \ref{2015xya5_bez_230}). For each VC, we measured the minimum comet-star distance, and we use a black dot when this distance was smaller than 1000~au, and a red dot otherwise. In this plot, there are 1079 black dots and 3921 red dots. The blue cross marks the nominal result. It is clear from  the $q$ histogram that the most probable value is a little smaller than the nominal one.

However, to observe the overall uncertainty of C/2015~XY1 orbit at 2.0 Myr ago, we had to add the effect of the stellar data uncertainty. The result of a calculation performed for 12\,000 pairs of a star clone and a comet clone acted by the former,  is presented in Fig. \ref{fig_combined_uncert}. Apart from black and red dots, we also use green dots in this plot, which corresponds to all pairs with the proximity distance greater than 2000~au. There are 1725 black, 5111 red, and 4324 green dots in this plot. For an easier comparison, we keep the same scales as in Fig. \ref{fig_comet_uncert} and as a result, 841 black dots from the right tail of the distribution are not visible. These absent dots are extremely spread in both elements. For $1/a$, the covered range is from -1431.32 to 38.78, so some small fraction of hyperbolic orbits were also obtained. The largest perihelion distance obtained in this simulation is 46\,160~au.

In Fig. \ref{fig_combined_uncert}, we observe that the most probable value of a perihelion distance is significantly smaller ($\sim$50~au) than the nominal $q=221$~au (blue cross). This effect results from a large number of cases when a star-comet minimum distance is larger than the nominal (all green dots and most of the red dots). This is well illustrated in the inset of Fig. \ref{P0506_scatter}.

In light of the above results, it should be obvious that tracing the motion of C/2015~XY1 further to the past, and calculating the results of the strong perturbation caused by P0230 2.8 Myr ago, seems useless since almost all physically acceptable results will be possible after taking into account P0230 uncertainties. For example, according to our current knowledge, we cannot say anything about  the previous perihelion distance of this comet -- even a hyperbolic past orbit is possible due to the strong interaction with P0506.

\section{Summary and conclusions}
\label{sec:conclusions}

Here, we propose a set of algorithms and methods for practical use in the Solar System small body dynamics investigation when Galactic and stellar perturbations have to be taken into account. The proposed approach allows for fast and precise numerical integration of the equations of motion formulated in the Solar System barycentric reference frame. The past and future motion of LPCs for millions years can be studied in detail and the reliability of the results can be effectively controlled using the Monte Carlo sampling of the input data.

The equations of motion derived in Section \ref{differential} were carefully tested by a comparison with the result of a quadruple precision calculation performed in the galactocentric frame. These equations give the full available precision. The numerical stellar ephemerides described in Section \ref{ephemeris} are optimized for the standard double precision calculations. Their application increases the precision by 6 significant digits, while reducing the computation time by a factor 100. There is no trade-off between the precision and speed -- both the quantities are improved.  These calculations are over 20\,000 times faster than their previous versions  performed in the quadruple precision, which is necessary to obtain a comparable precision in the galactocentric frame.
The numerical ephemerides of a set of potential stellar perturbers, as well as the example C code for their application and the approximate polynomial galactocentric Sun ephemerides are publicly available.

To present possible applications of the proposed methods, we performed the overall tests of the stellar perturbations effect on all LPCs included in the CODE database. We show which stars from our current list are the most effective comet motion perturbers. We also present a detailed analysis of one particular example -- the past motion of comet C/2015~XY1. This comet orbit was strongly perturbed as a consequence of a very close passage of the star P0506 about 1.8 Myr ago, but it was also perturbed at an earlier time by P0230. We described the influence of both comet and stellar parameter uncertainties on the possibility to obtain this comet past dynamic state.

\begin{acknowledgements}
 We are very grateful to the reviewer, Marc Fouchard, for his insightful comments that have helped us improve this manuscript. This research has made use of the SIMBAD database, operated at CDS, Strasbourg, France. This research has also made use of the VizieR catalog access tool, CDS, Strasbourg, France (DOI: 10.26093/cds/vizier). This work has made use of data from the European Space Agency (ESA) mission {\it Gaia} (\url{https://www.cosmos.esa.int/gaia}), processed by the {\it Gaia} Data Processing and Analysis Consortium (DPAC, \url{https://www.cosmos.esa.int/web/gaia/dpac/consortium}). Funding for the DPAC has been provided by national institutions, in particular the institutions participating in the {\it Gaia} Multilateral Agreement. The calculations which led to this work were partially performed with the support from the project  "GAVIP-GC: processing resources for Gaia data analysis" funded by European Space Agency (4000120180/17/NL/CBi).
\end{acknowledgements}

\bibliographystyle{aa} 
\bibliography{PAD30}

\appendix
\section{Heliocentric accelerations}
\label{Ap:der}

The transformation of equations of motion to a form that does not require subtraction of almost equal terms dates back to the Encke's method of numerical integration, where the relative acceleration with respect to the nominal Keplerian orbit had to be computed \citep{BroCle:1961}. Here, the context is different and the potentials are more complicated, but the spirit and some of the details remain similar.

For example, transforming $\vec{f}_b$, initially defined through Eq.~(\ref{e13}), we proceed as follows. First, the coefficients of $\vec{r}$ and $\vec{R}_\odot$ are separated:
\begin{equation}
  \frac{\vec{R}}{\left(R^2+b_b^2\right)^\frac{3}{2}}-\frac{\vec{R}_\odot}{\left(R_\odot^2+b_b^2\right)^\frac{3}{2}}=
   \frac{1}{q_1^3}\left(\frac{p_1^3-q_1^3}{p_1^3}\, \vec{R}_\odot + \vec{r}\right),
\end{equation}
where $p_1$ and $q_1$, given by Eqs.~(\ref{p1:def},\ref{q1:def}), are almost equal. Then, the problematic term is converted into
\begin{equation}
    \frac{p_1^3-q_1^3}{p_1^3} = \frac{p_1^6-q_1^6}{p_1^3 \left(p_1^3+q_1^3\right)}.
\end{equation}
Further,
\begin{equation}
     p_1^6-q_1^6 = \left(p_1^2-q_1^2\right)\left(p_1^4 + p_1^2 q_1^2 + q_1^4\right),
\end{equation}
and the first factor is neatly transformed into
\begin{eqnarray}
 p_1^2-q_1^2 & = & \left(R^2+b_b^2\right)-\left(R_\odot^2+b_b^2\right) \nonumber \\
   &=& \left(R_\odot^2 + 2 \vec{r} \cdot \vec{R}_\odot + r^2\right) - R_\odot^2 \nonumber\\
   &=&  2 \vec{r} \cdot \vec{R}_\odot + r^2.
\end{eqnarray}
Even if the two components in the last line have different signs, they are far from having the same order of magnitude. The same device
has been applied for the $\vec{f}_d$.

The treatment of $\vec{f}_h$ is also simple. The coefficient of $\vec{R}_\odot$ involves
\begin{equation}
  1-  \frac{\left(a_h + R\right) R }{\left(a_h + R_\odot\right) R_\odot} =
  \frac{a_h \left(R_\odot - R\right) + R_\odot^2- R^2 }{\left(a_h + R_\odot\right) R_\odot},
\end{equation}
where
\begin{equation}
    R_\odot - R  = \frac{ R_\odot^2- R^2 }{R_\odot+R},
\end{equation}
and
\begin{equation}\label{clue}
     R_\odot^2- R^2 =  - 2 \vec{r} \cdot \vec{R}_\odot - r^2.
\end{equation}

In the NFW model of the halo potential, one meets a new problem of subtracting logarithms with almost equal arguments or, equivalently, of
a logarithm whose argument is a fraction with almost equal numerator and denominator. This obstacle is bypassed by the conversion:
\begin{eqnarray}
  \ln \frac{a_H+R_\odot}{a_H+R} &=&  \ln \frac{\left[ \left(a_H+R\right)+\left(R_\odot -R\right) \right]\left( R+R_\odot \right)}{\left(a_H+R\right)\left( R+R_\odot \right)} \nonumber\\
  &=&  \ln \left[ 1 + \frac{ R_\odot^2- R^2}{\left(a_H+R\right)\left( R+R_\odot \right)}\right],
\end{eqnarray}
and recalling Eq.~(\ref{clue}), we find the form $\ln(1+\xi)$, where $|\xi| \ll 1$ does not involve a problematic subtraction. We note, however, that
direct application of the natural logarithm function, implemented in standard compilers, may also degrade the accuracy. For small values of $\xi,$ it is recommended to
use specially designed functions like \texttt{log1p} in C/C++, or to resort to the power series expansion.

\section{Polynomial ephemeris of the Sun\label{app-ephem}}

For the cases where only the Galactic tides are to be taken into account in a  small body motion in the Solar System,  it is sufficient to have only the Sun's galactocentric positions calculated in advance while using the formulae proposed in Sect.~\ref{differential}. The simplest and fasted way is to use polynomial approximations of these positions obtained on the basis of the precise numerical integration. In this appendix, we propose such ephemerides, both for backward and forward small body dynamics studies. The coefficients of the fitted polynomials (all of them of the fifth degree) are presented in Tables \ref{tab-sol-eph-back} and \ref{tab-sol-eph-forth}. In these tables, ephemerides for the past and future 30~Myr of the Sun's galactocentric motion are presented. In both cases the whole interval is divided into six parts to ensure the similar accuracy of the whole ephemeris. The maximum difference between numerical integration results and positions obtained with these ephemerides is below $1\times10^{-4}$ parsecs. It is worth recalling here that the distance of the Sun from the Galactic center is known with the uncertainty of hundreds of parsecs and for the vertical Sun position only the sign (positive) is certain. This makes the proposed ephemerides accuracy quite satisfactory.

The structure of both tables is identical: we present six polynomial segments covering the whole 30~Myr intervals. Each segment is described in the appropriate table with four rows: the start and the end of the validity time interval expressed in Myr, and three rows with the polynomial coefficients for $X_\odot$, $Y_\odot,$ and $Z_\odot$ galactocentric positions of the Sun.

If we denote the coefficients in any row as $a_0, a_1, a_2,$ \dots $, a_5$, then each coordinate in each interval can be obtained from the following formula:
\begin{equation}
  a(t_0+t) = (((((a_5 t)+a_4) t +a_3) t+a_2) t + a_1) t + a_0,
\end{equation}
where for greater speed and better precision, we used the Horner scheme. In this formula, $t_0$ is the initial epoch of the time interval of interest, $t$ is the time from the beginning of the interval to the point of interest, and $a$ represent any of the $X_\odot$, $Y_\odot$ or $Z_\odot$ Galactocentric coordinate of the Sun at the epoch $(t_0+t)$.

It should be noted that both these ephemerides are obtained by polynomial least square fitting to the results from a precise numerical integration of the Sun and all 407 stars, based on a list from the StePPeD database but updated to the {\it Gaia} EDR3 results. All mutual interactions between stars were taken into account. As a result, the division of the whole intervals of 30 Myr is different and unequal because it depends on these mutual interactions during star-star encounters or star-sun interactions.
We note that the computer readable files with these ephemerides are publicly available at the StePPeD database internet site.

\clearpage
\begin{sidewaystable}
\caption{Polynomial backward ephemeris of the Sun; see text for a detailed description. \label{tab-sol-eph-back} }
\centering
{\footnotesize{ 
\begin{tabular}{rrrrrr}
\multicolumn{1}{l}{time interval:} &   \multicolumn{1}{l}{0.000000000000000}  &  \multicolumn{1}{l}{$-$2.449311977543401}  & & & \\
$-$8.400000002057879e+03 &  1.135181550880720e+01 &  3.643862347504248e+00  &   1.103925401461188e$-$04  &  $-$1.084953291607976e$-$03  &  $-$1.275222673372003e$-$04 \\
4.587283194723568e$-$05 &  2.600121512460876e+02 &  1.767733474010741e$-$03  &  $-$3.743792832101693e$-$02  &  $-$4.602068851860490e$-$04  &  $-$1.096933592070080e$-$04 \\
1.700001208490988e+01 &  7.414611785275893e+00 & $-$5.118640113957033e$-$02  &  $-$1.071726403376998e$-$02  &  $-$1.676934039001960e$-$03  &  $-$2.634239541292695e$-$04 \\
\multicolumn{1}{l}{time interval:} &
\multicolumn{1}{l}{ $-$2.449311977543401}  &  \multicolumn{1}{l}{$-$2.794718679721595}  & & & \\
$-$8.051110411270996e+03 &  6.777397531207872e+02 &  5.125092546698994e+02  &   1.941898576821398e+02  &   3.703277405464308e+01  &   2.823540812434333e+00 \\
$-$1.966861038857007e+01 &  2.222683177982121e+02 & $-$2.896215624694570e+01  &  $-$1.114699343493198e+01  &  $-$2.130210858615920e+00  &  $-$1.633363313309615e$-$01 \\
1.557200379958330e+01 &  4.739911781896996e+00 & $-$2.048501264493539e+00  &  $-$7.531779534890005e$-$01  &  $-$1.388351137604125e$-$01  &  $-$1.031076410323967e$-$02 \\
\multicolumn{1}{l}{time interval:} &  \multicolumn{1}{l}{$-$2.794718679721595}  &  \multicolumn{1}{l}{$-$8.798809561544877}  & & & \\
$-$8.399962783848834e+03 &  1.137416492246366e+01 &  3.650769032948845e+00  &   2.826690360530395e$-$03  &  $-$2.314504765501216e$-$04  &   1.798622734510769e$-$06 \\
2.591994029216949e$-$02 &  2.600238260984271e+02 &  1.838663878152236e$-$03  &  $-$3.741806009003032e$-$02  &  $-$4.787599636650243e$-$05  &   2.277646269310953e$-$06 \\
1.702581846927255e+01 &  7.428797634906145e+00 & $-$4.535903368534015e$-$02  &  $-$6.350314736788380e$-$03  &   1.100201173122527e$-$04  &   6.243124803047839e$-$06 \\
\multicolumn{1}{l}{time interval:} &  \multicolumn{1}{l}{$-$8.798809561544877}  & \multicolumn{1}{l}{$-$15.135043351089207}  & & & \\
$-$8.399606225887133e+03 &  1.151289365887244e+01 &  3.669408099936796e+00  &   3.580144862441055e$-$03  &  $-$2.651606688743473e$-$04  &  $-$7.878525003636557e$-$07 \\
5.481570775850741e$-$01 &  2.602651112714771e+02 &  4.550644828002848e$-$02  &  $-$3.355757901653256e$-$02  &   1.188524427064714e$-$04  &   5.113058007725760e$-$06 \\
1.706057250939377e+01 &  7.432865128379767e+00 & $-$4.917955923895045e$-$02  &  $-$7.537486101964606e$-$03  &  $-$1.760414617943653e$-$05  &   1.443519176397130e$-$06 \\
\multicolumn{1}{l}{time interval:} & \multicolumn{1}{l}{$-$15.135043351089207}  & \multicolumn{1}{l}{$-$26.102447861828644}  & & & \\
$-$8.399346405086311e+03 &  1.151815504011466e+01 &  3.659876581585557e+00  &   2.326388115975922e$-$03  &  $-$3.245702429246902e$-$04  &  $-$1.762087178661271e$-$06 \\
$-$1.071419805472942e+00 &  2.596870424177994e+02 & $-$4.278517823090100e$-$02  &  $-$4.061911400140923e$-$02  &  $-$1.707597211048486e$-$04  &   3.155774154345478e$-$07 \\
1.618160824278573e+01 &  7.118837684149699e+00 & $-$9.490043708660925e$-$02  &  $-$1.090162366926758e$-$02  &  $-$1.419092538587968e$-$04  &  $-$3.931166083283932e$-$07 \\
\multicolumn{1}{l}{time interval:} & \multicolumn{1}{l}{$-$26.102447861828644}  & \multicolumn{1}{l}{$-$30.000000000000000}  & & & \\
$-$7.990527280582611e+03 &  8.459635463987752e+01 &  8.882764149264744e+00  &   1.888879675399864e$-$01  &   3.006144187470992e$-$03  &   2.201529892485608e$-$05 \\
$-$8.249915678630405e+02 &  1.123784885697437e+02 & $-$1.057307230153560e+01  &  $-$4.168438366543535e$-$01  &  $-$6.889196506322961e$-$03  &  $-$4.765953135696503e$-$05 \\
5.689216937933580e+00 &  5.229966950574662e+00 & $-$2.311606775528041e$-$01  &  $-$1.582814936307915e$-$02  &  $-$2.312436670503466e$-$04  &  $-$1.043609629104035e$-$06 \\
\end{tabular}
}}      
\end{sidewaystable}

\begin{sidewaystable}
\caption{Polynomial forward ephemeris of the Sun; see text for a detailed description. \label{tab-sol-eph-forth} }
\centering
{\footnotesize{ 
\begin{tabular}{rrrrrr}
\multicolumn{1}{l}{time interval:} &   \multicolumn{1}{l}{0.000000000000000}  &  \multicolumn{1}{l}{1.479496969724100}  & & & \\
$-$8.400000009959480e+03  &   1.135237349801379e+01 &  3.641491931630962e+00  &   9.904794331577099e$-$03  &  $-$7.825751370654431e$-$03  &   2.469899964808171e$-$03       \\
4.048151572533243e$-$05  &   2.600098443265476e+02 &  5.651697270344758e$-$03  &  $-$5.206212410645247e$-$02  &   1.426674531053425e$-$02  &  $-$4.866323571788590e$-$03   \\
1.699997817913812e+01  &   7.415359175909785e+00 & $-$5.534940854948545e$-$02  &   5.545061890083549e$-$03  &  $-$1.247015940106314e$-$02  &   4.208156952482975e$-$03     \\
\multicolumn{1}{l}{time interval:} &   \multicolumn{1}{l}{1.479496969724100}    &   \multicolumn{1}{l}{5.862181103455256} & & & \\
$-$8.399999565454938e+03  &   1.134887175007549e+01 &  3.647801309337811e+00  &   1.344931739571749e$-$03  &  $-$2.946396654956642e$-$04  &  $-$1.275320958418031e$-$06     \\
9.936985863532755e$-$03  &   2.600007167848327e+02 &  2.659993834643992e$-$03  &  $-$3.856766092844950e$-$02  &   9.181411922605127e$-$05  &  $-$4.975121368183102e$-$06   \\
1.699033897593918e+01  &   7.426446726929897e+00 & $-$5.552561724746034e$-$02  &  $-$5.364423768157968e$-$03  &  $-$2.564128445427606e$-$04  &   1.813511848726224e$-$05   \\
\multicolumn{1}{l}{time interval:} &   \multicolumn{1}{l}{5.862181103455256}    &   \multicolumn{1}{l}{6.881513553048107} & & & \\
$-$8.133445740021927e+03  &  $-$1.986657868342002e+02 &  6.978901127395488e+01  &  $-$1.040539905932462e+01  &   8.177004127832183e$-$01  &  $-$2.569783202358540e$-$02     \\
5.916194831792222e+01  &   2.130282232323002e+02 &  1.490621384452508e+01  &  $-$2.400651300639794e+00  &   1.871153233584631e$-$01  &  $-$5.922736423299913e$-$03         \\
2.270188552553307e+01  &   3.232135866404568e+00 &  1.154742448172056e+00  &  $-$1.774690832759001e$-$01  &   1.184009666594990e$-$02  &  $-$3.182803118830619e$-$04       \\
\multicolumn{1}{l}{time interval:} &   \multicolumn{1}{l}{6.881513553048107}    &  \multicolumn{1}{l}{13.718717932634034} & & & \\
$-$8.400011646849596e+03  &   1.139624914369730e+01 &  3.632422012392106e+00  &   3.253087161664779e$-$03  &  $-$4.155963407616601e$-$04  &   1.832628238080319e$-$06       \\
1.303466117960852e$-$01  &   2.599261970248300e+02 &  1.717877829573644e$-$02  &  $-$3.932625051232038e$-$02  &   3.056825001898819e$-$05  &   1.272961983812298e$-$06     \\
1.688465793791785e+01  &   7.451248955385243e+00 & $-$5.404027912658385e$-$02  &  $-$6.849944503754613e$-$03  &   1.422432012482767e$-$05  &   2.136369109298331e$-$06     \\
\multicolumn{1}{l}{time interval:} &  \multicolumn{1}{l}{13.718717932634034}    &  \multicolumn{1}{l}{19.550877368549660} & & & \\
$-$8.397023969392378e+03  &   1.050410808239803e+01 &  3.738881949388912e+00  &  $-$3.170536162029228e$-$03  &  $-$2.161759041534958e$-$04  &  $-$7.668710780866691e$-$07   \\
4.194976468254026e+00  &   2.587416992845421e+02 &  1.523846319842710e$-$01  &  $-$4.677713714157145e$-$02  &   2.227881421599462e$-$04  &  $-$4.393579419801620e$-$07     \\
1.706748864906339e+01  &   7.339021850508717e+00 & $-$3.082828543767888e$-$02  &  $-$9.131341773441770e$-$03  &   1.246357013286258e$-$04  &   1.201798398863710e$-$08     \\
\multicolumn{1}{l}{time interval:} &  \multicolumn{1}{l}{19.550877368549660}    &  \multicolumn{1}{l}{30.000000000000000} & & & \\
$-$8.402630649708150e+03  &   1.205686726742499e+01 &  3.565246755182795e+00  &   6.614330841878433e$-$03  &  $-$4.936426869819803e$-$04  &   2.396122933663810e$-$06       \\
5.169917913861354e+00  &   2.587808405179473e+02 &  1.185319479705065e$-$01  &  $-$4.347506987694289e$-$02  &   9.637370970754560e$-$05  &   1.308558354838703e$-$06       \\
2.028683641389472e+01  &   6.600590101621695e+00 &  3.724220377612755e$-$02  &  $-$1.228496486850992e$-$02  &   1.980826145723392e$-$04  &  $-$6.759264608864593e$-$07     \\
                        \end{tabular}
        }}      
\end{sidewaystable}

\clearpage

\end{document}